\documentclass[aps,pre,showpacs,twocolumn]{revtex4}
\usepackage{amsmath}
\usepackage{amssymb}
\usepackage{graphicx}
\begin{document}

\title{Array enhanced stochastic resonance and spatially correlated noises}

\author{P. F. G\'ora}
\email{gora@if.uj.edu.pl}

\affiliation{M.~Smoluchowski Institute of Physics and Complex Systems 
Research Center, Jagellonian University, Reymonta~4, 30--059~Krak\'ow, Poland}

\date{\today}

\begin{abstract}
We discuss the role of spatial correlations of the noise in the array 
enhanced stochastic resonance. We show numerically that the noises with 
negative correlations between different sites lead to significantly larger
values of the signal-to-noise ratio than the uncorrelated noises or noises
with positive correlations. If the noise is global, the system displays
only the conventional stochastic resonance, without any array enhancement.
\end{abstract}


\maketitle

\section{Introduction}

Stochastic resonance (SR) \cite{SR} is the best-known example of the
constructive role of noise. The SR is a phenomenon in which the response
of a~system is optimized by the presence of a specific level of noise.
It has been detected in so many seemingly different systems that it has been
claimed to be ``an inherent property of rate-modulated series of events'' 
\cite{Bezrukov1}. It has been recently suggested that the functioning of 
important natural devices, e.g., communication and information
processing in neural systems or subthreshold signal detection
in biological receptors, rely on phase synchronization rather
than stochastic resonance \cite{Freund}, but this does not exclude
the possibility that some natural devices may rely on the SR
or that effective artificial detectors that use this feature may be
constructed and operated.

It has been observed that the SR gets enhanced if an array of similar nonlinear
elements collectively responds to the same signal. This phenomenon
has been termed the array enhanced stochastic resonance (AESR). It was first
observed in chains of nonlinear oscillators \cite{Bulsara} and later, mostly
without explicitly using the term AESR, in 
arrays of FritzHugh-Nagumo model neurons \cite{Collins}, in ion channels
\cite{Bezrukov3}, in ensembles of nondynamical elements with internal
noise independent on the incoming signal or modulated by it \cite{Collins2},
in multilevel threshold systems \cite{multilevel}, in a globally coupled
network of Hodgkin-Huxley model neurons \cite{Bambi}, and, recently, in
arrays of Josephson junctions \cite{prb}. The experimental work on detecting
the SR in the mammalian brain tissue \cite{brain} should also be mentioned
as a whole \textit{array} of neurons was stimulated. The accumulated knowledge
of systems exhibiting the AESR is now considerable, but there are still
important issues that need to be clarified. We want to address two such
points in the present paper.

First, in nearly all papers mentioned above, the specific dynamics --- the
``internal'' dynamics of systems such as nonlinear oscillators or 
model neurons, or dynamical coupling between various elements
--- played an important role. It is thus not entirely clear what features
of the AESR are generic, and what depends on the details of interactions.
We will discuss the AESR in arrays of nondynamical threshold elements.
Such elements, first introduced in Ref.~\cite{nondynamical1} and later
discussed in Refs.~\cite{nondynamical2}, are known to display the SR
when subjected to a~periodic subthreshold signal with an additive 
noise. The AESR, if indeed present in such an array, is not obscured by any 
effects resulting from a specific dynamical model, and only its generic
features should show up. The interesting research presented in 
Ref.~\cite{Collins2} already considered the AESR in ensembles of nondynamical 
elements, but we want to re-examine this problem because of the special 
treatment of noises in Ref.~\cite{Collins2} --- see a discussion below.

Second, the role of spatial correlations between the noises has been almost
neglected in the
existing research on the AESR. The noises discussed were local and uncorrelated
in Refs.~\cite{Bulsara,Collins,Bezrukov3,multilevel,prb}, global in 
Ref.~\cite{brain}, and in the form of a mixture of uncorrelated local and global 
noises in Refs.~\cite{Collins2,Bambi}. As correlations between various noises 
can result in very interesting physical phenomena, ranging from drastic changes
in activation rates \cite{Telejko,duffing}, through current reversal in 
Brownian ratchets \cite{Luczka}, to an effective nullifying one of the
noises \cite{singh}, and many others, it is interesting to examine their
impact on the AESR. The authors of Ref.~\cite{Bambi} claimed
that spatial correlations between the noises weakened the response of an array
of neurons, but the spatial correlations discussed there were induced by
the presence of a global noise. We will show that spatial correlations between
\textit{local} noises can have a constructive effect.

This paper is organized as follows: In Section~II we present numerical results
for arrays of uncoupled threshold elements. In Section~III we discuss the role
of spatial correlations between the noises in a chain of nonlinear oscillators.
Concluding remarks are given in Section~IV.

\section{Uncoupled threshold elements}

A single threshold device fires whenever the signal on its input exceeds
the threshold. Here we consider 
an array of $N$ such devices, acting in parallel in response to a common signal; 
the average (or integrated) output of all individual devices is taken as the output 
of the whole array:

\begin{subequations}\label{aesr:array}
\begin{eqnarray}
g_i(t) &=&
\begin{cases}
1&\mathrm{if\ }A\sin(\omega t+\phi) + \eta_i(t) \geqslant 1\\
0&\mathrm{otherwise}
\end{cases}
\\
g(t) &=& \frac{1}{N}\sum\limits_{i=1}^{N} g_i(t)\,.
\end{eqnarray}
\end{subequations}

\noindent Here $\omega$ is the frequency of the signal (we take $\omega=2\pi$), 
$\phi$ is the initial phase,
and $A$ is the amplitude; we take $A=0.8$ to make the signal subthreshold.
$g(t)$ is the output of the array and $\eta_i(t)$ are the noises. We take 
them to be zero-mean Gaussian white noises (GWNs), possibly spatially 
correlated:

\begin{equation}\label{aesr:etacorrelated}
\left\langle\eta_i(t)\right\rangle=0\,,\quad
\left\langle\eta_i(t)\eta_j(t^\prime)\right\rangle
=\sigma^2 C_{ij}\,\delta(t-t^\prime)\,.
\end{equation}

\noindent The matrix $\mathbf{C}=\left[C_{ij}\right]$ represents spatial
correlations of the noises. It must be symmetric and positively definite.
$\sigma$ is a parameter controlling the intensity of the noises. This system
is similar to, but different from that discussed in the first part of 
Ref.~\cite{Collins2},
where the local noise was added to the output of each element \textit{after} 
the element had decided whether to fire in a response to the signal
contaminated by a global noise. Thus, the output of a single element was
not binary. Rather than that, it could assume, in principle, any value,
positive or negative. Moreover, as all elements received identical inputs,
they all fired or not fired in unison. After an appropriate scaling, the 
collective output of the whole array was equivalent to a binary series
contaminated by a GWN. The local noises in Ref.~\cite{Collins2}
were, by assumption, spatially uncorrelated. In our approach, the internal
noise is added to the signal at each site \textit{before} the threshold
elements make their decisions whether to fire or not. The internal noise
represents fluctuations in the connections or environmentally induced
random perturbations, but the output of every threshold element remains
binary.

\begin{figure}
\includegraphics[scale=0.67]{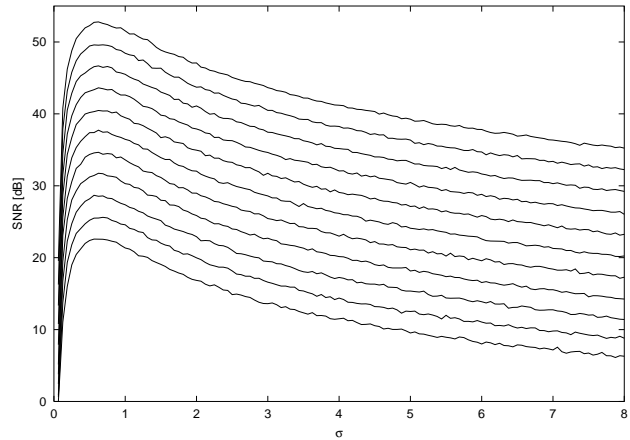}
\caption{The AESR for the nondynamical system \eqref{aesr:array}. 
The curves correspond, bottom
to top, to the arrays of lengths 1, 2, 4, 8, 16, 32, 64, 128, 256, 512, and
1024, respectively. The noises acting on different elements of the arrays 
are not correlated, $C_{ij}=\delta_{ij}$.}
\label{aesr:fig-aesr}
\end{figure}

We also start with uncorrelated, or local, noises, $C_{ij}=\delta_{ij}$.
We calculate the signal and the noises with a time step $h=1/32$ and calculate
the output of the nondynamical system \eqref{aesr:array}.
We use Marsaglia algorithm \cite{Marsaglia} to generate
the GWNs; we use the famous Mersenne Twister \cite{Mersenne} as the underlying
uniform generator. For each array, we collect a time series of 4096
elements, calculate its power spectrum and calculate the signal-to-noise
ratio (SNR):

\begin{equation}\label{aesr:SNR}
\mathrm{SNR} = 10\,\log_{10}
\frac{\mathrm{power\ density\ at\ the\ signal\ frequency}}
{\mathrm{background\ power\ density}}\,.
\end{equation}

\noindent For each $N$ and $c_1$ we average the results
over 512 realizations of the noises and initial phases of the signal.
Final results are presented in Fig.~\ref{aesr:fig-aesr} and their
interpretation is clear: The SNR increases significantly for all noise strengths
as the array size doubles. We can see that the AESR is not a result of any specific
dynamics but is present also in arrays of uncoupled elements. This, together
with previous results for coupled systems, shows that the AESR is a generic feature
of arrays of elements that individually display the SR.

Note that we have observed the AESR for a \textit{local} noise. If the noise were
\textit{global}, $\forall\,i,j:\ \eta_i(t)\equiv\eta_j(t)$ (or $C_{ij}\equiv1$),
no array enhancement would be possible. Indeed, for a global noise, each element
of the array receives identical input, and the collective output of the whole array 
is identical with that of a single element. The array does not display the AESR,
but only the conventional SR. This simple argument shows that it is
the differences between the noises at different sites that cause the AESR. This
observation leads to the question of the role of spatial correlation between the
noises. Qualitatively speaking, noises with large correlations are ``nearly
identical'' and the enhancement of the SR should be small. The enhancement of the SR
should increase as the correlations decrease. We will now see that this is indeed
the case.

Let $\mathbf{C}=\mathbf{GG}^T$ be the Cholesky decomposition \cite{Golub} of 
the correlation matrix $\mathbf{C}$ and let $\widetilde{\boldsymbol\eta}(t)$
be a vector of spatially uncorrelated, zero mean GWNs, 
$\left\langle\widetilde{\eta}_i(t)
\widetilde{\eta}_j(t^\prime)\right\rangle=\sigma^2\delta_{ij}\,\delta(t-t^\prime)$.
Then $\boldsymbol\eta=\mathbf{G}\widetilde{\boldsymbol\eta}$ has correlations
of the form \eqref{aesr:etacorrelated}. In the following, we will consider only
correlations between the nearest neighbors in the array of threshold elements.
Specifically,

\begin{equation}\label{aesr:C}
C_{ij}=
\begin{cases}
1&\mathrm{if\ }i=j\,,\\
c_1&\mathrm{if\ }|i-j|=1\,,\\
0&\mathrm{otherwise}\,,
\end{cases}
\end{equation}

\noindent where $|c_1|\leqslant0.5$ in order to keep the matrix $\mathbf{C}$
positively definite. For each specific value of $c_1$, the Cholesky
decomposition needs to be performed only once; later, during the simulation,
vectors of independent Gaussian variables are generated and multiplied
by the Cholesky factor $\mathbf{G}$. Since for a tridiagonal correlation
matrix $\mathbf{C}$, the Cholesky factor has only two non-zero elements per
row, generating GWNs with nearest-neighbor correlations is only twice as
computationally expensive as generating independent GWNs, or even less than that
given the fact that generating the independent Gaussian variables is the
most computationally intensive part of the procedure.

\begin{figure}
\includegraphics[scale=0.67]{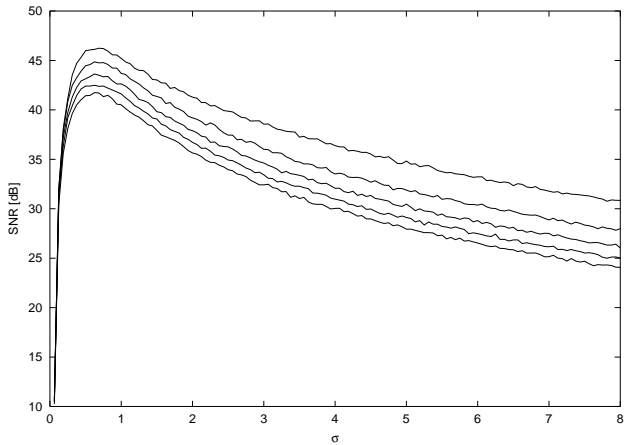}
\caption{The effect of spatial correlations on the AESR. 
The curves correspond, top to bottom, to $c_1=-1/2$, $-1/4$, $0$, $1/4$, and $1/2$,
respectively. The array consists of 128 threshold devices.}
\label{aesr:fig-near}
\end{figure}

We now simulate the system
\eqref{aesr:array} in the manner described above for various values of $c_1$.
Results for an array of $N=128$ elements are plotted in Fig.~\ref{aesr:fig-near}.
We can see that noises with positive spatial correlations lead to a smaller 
enhancement of SR than the independent noises. On the other hand, the SNR 
increases as $c_1$ becomes negative and approaches $-0.5$. The effect of 
negative correlations (anticorrelations) becomes larger as the noise strength 
increases. Similar results have been observed for arrays of different lengths.

\begin{figure}
\includegraphics[scale=0.74]{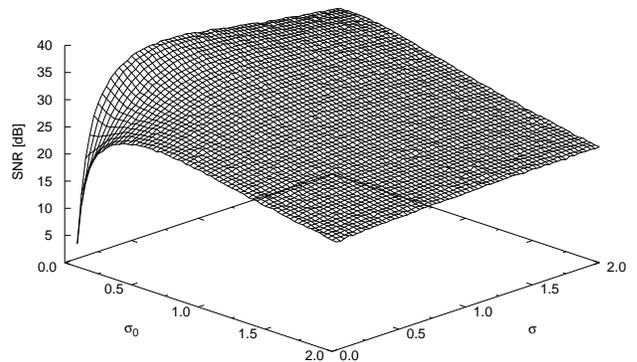}
\caption{The AESR as a function of the strength of the global (external)
noise $\sigma_0$ and the strength $\sigma$ of the additional noise applied 
to the array. The additional noise has spatial correlations in the form
\eqref{aesr:C} with $c_1=-0.5$, and the array size $N=8$.}
\label{aesr:fig-addnoise}
\end{figure}

Clean signals are rare in nature.
Suppose that the incoming signal is contaminated by a noise that
cannot be controlled. To improve detection of this signal, we pass it 
through an array of threshold elements and we apply additional noise
that we can control to each of the elements. The external noise acts here
as a global noise; the additional (local) noise is not correlated to the 
global noise. The above discussion suggests that the additional noise should 
have negative spatial correlations. If the global noise is weak (below the 
peak of the ordinary SR), the additional anticorrelated noise
can significantly enhance the SNR, Fig.~\ref{aesr:fig-addnoise}. However,
if the global noise is large, the enhancement provided by the array is only
marginal. This is similar to the result reported in Ref.~\cite{Bambi}, 
where the presence of a~strong global noise markedly deteriorated the
performance of the system studied. 

\begin{figure*}
\includegraphics{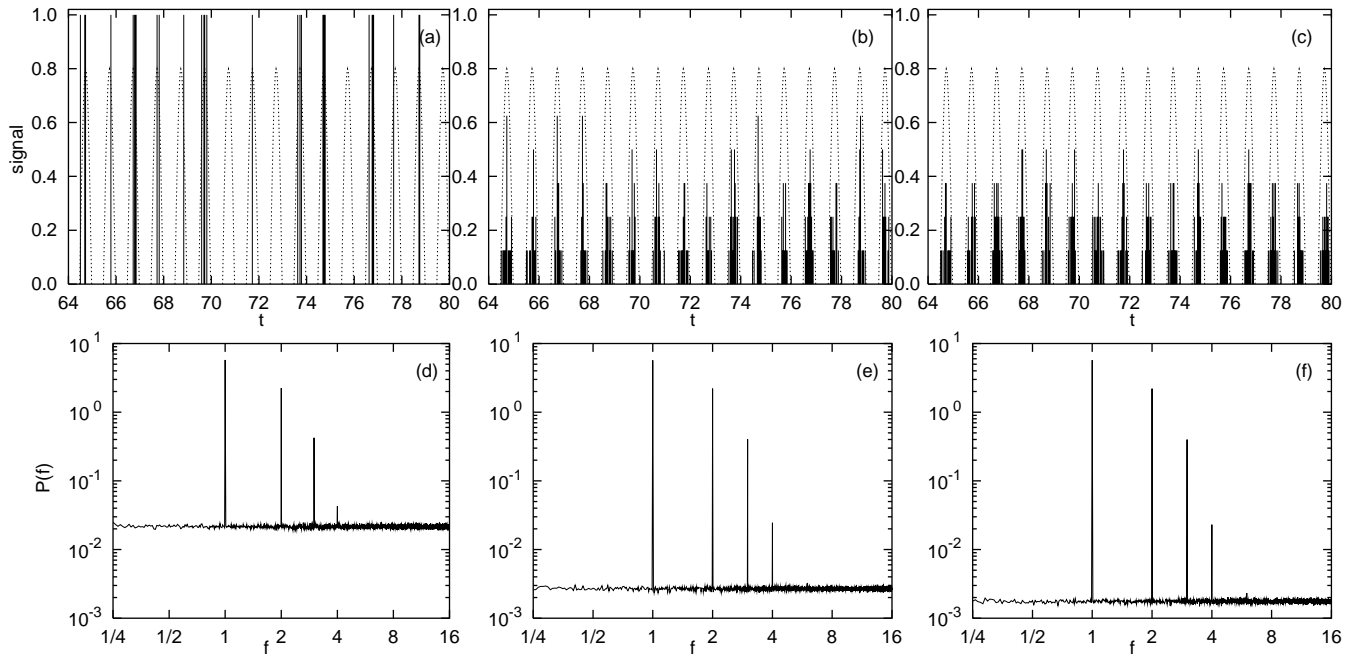}
\caption{Upper row: typical realizations of signals outgoing from a single 
threshold device (panel a), from an array of 8~devices with independent
noises (panel b), and from a similar array with anticorrelated noises (panel c). 
Positive values of the incoming periodic signal are shown by a broken line. 
Bottom row: Power spectra averaged over 512 realizations of the noise. 
Noise parameters are the same as in the corresponding panels in the upper row. 
In all cases $\sigma=0.375$.}\label{aesr:fig-panels}
\end{figure*}

It is important to understand the ``microscopic'' mechanism responsible for
the AESR. In the conventional SR, the threshold element may occasionally
misfire or miss some peaks of the signal, cf.\ Fig.~\ref{aesr:fig-panels}a. 
Also the shape of the incoming signal is not resolved by the outgoing 
binary signal. When an array of such elements acts in parallel, if one of
the elements makes a mistake (fires when the signal is low or fails to fire
when the signal is strong), other elements that are not positively correlated
with it are not likely to repeat the mistake (Figs.~\ref{aesr:fig-panels}b and~c). 
This effect is even stronger
when the other elements are negatively correlated with the one that makes
the mistake: negative correlations between noises at different sites tend
to correct the mistakes, while positive correlations tend to repeat them. 
As a result, the shape of the incoming signal is resolved much better.
This does not lead to a visible increase in the height of the signal
peak in the power spectrum, but it does lead to a significant decrease
of the noise background, Fig.~\ref{aesr:fig-panels}, bottom row. Thus,  
lowering of the flat noise background is primarily responsible for the 
increase in the SNR. 

On a more formal level, let $p$ be the probability that a single element
fires. This probability depends on the current phase of the incoming signal, 
on the signal's amplitude, and on the noise level. If exactly $k$ out of 
$N$ elements fire, the array's output equals $g=k/N$. If the noises at 
different sites are mutually independent (uncorrelated), the probability 
of such an event is given by the binomial distribution:

\begin{subequations}\label{aesr:binomial}
\begin{equation}\label{aesr:binomial1}
P_N\bigl(g(t)=k/N\bigr)
=\binom{N}{k} p^k(1-p)^{N-k}\,.
\end{equation}

\noindent In an array twice as large with other parameters
the same, exactly $2k$ elements should fire to produce the same output.
The probability of this event is

\begin{equation}\label{aesr:binomial2}
P_{2N}\bigl(g(t)=k/N\bigr)
=\binom{2N}{2k} p^{2k}(1-p)^{2(N-k)}\,.
\end{equation}

\end{subequations}

\noindent The distributions \eqref{aesr:binomial} have the same expectation
values $\bar k/N$, but for all values of $0<p<1$ the distribution 
\eqref{aesr:binomial2} is narrower than the distribution 
\eqref{aesr:binomial1}. More importantly, the distribution 
\eqref{aesr:binomial2}, corresponding to the larger array, allows for
a~more dense output, with values $(\bar k\pm\frac{1}{2})/N$ being
more probable than $(\bar k\pm1)/N$ etc. Consequently, in larger arrays,
wildly ``wrong'' outputs are less probable. This leads to lowering of the
noise background and to an increase of the SNR. 

If the noises are spatially correlated, the probability that exactly $k$ 
elements fire is no longer given by the binomial distribution. We have
not been able to derive an exact formula for this probability, but the general
mechanism of the SNR increase with the array size appears to be similar to
that for the uncorrelated noises. Note that for the case presented in 
Fig.~\ref{aesr:fig-panels}, introducing the maximal negative correlations
between the nearest neighbors lowers the background by a factor of the order
of~2, increasing the SNR by $10\log_{10}2\simeq3$~dB, or about 10\% of 
the total.

\section{A coupled system}

\begin{figure*}
\includegraphics{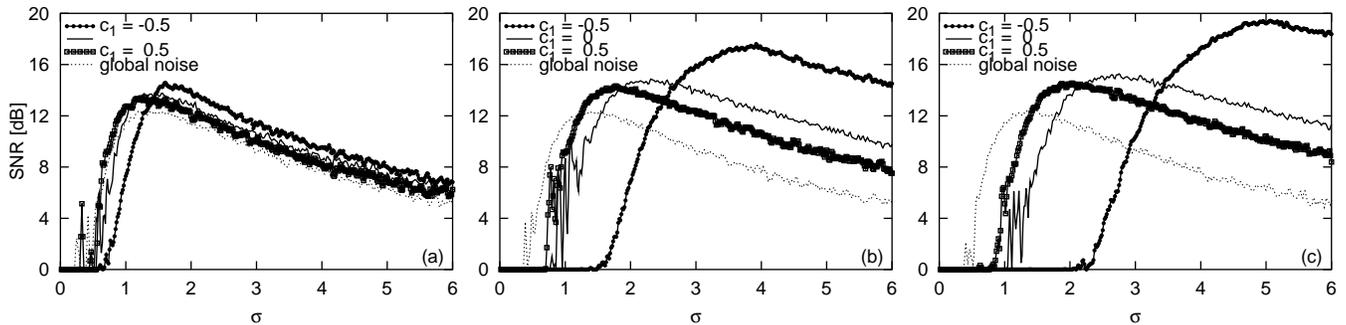}
\caption{Stochastic resonance for the coupled system
\eqref{aesr:coupled} with a global noise, a maximally correlated
noise ($c_1=1/2$), an uncorrelated (local) noise ($c_1=0$),
and a maximally anticorrelated noise ($c_1=-1/2$). Panel
(a) $\varepsilon=1$, panel (b) $\varepsilon=8$, panel (c)
$\varepsilon=16$.}\label{aesr:fig-coupled}
\end{figure*}

In order to verify whether similar effects are present in
coupled systems, we consider the same system that was discussed
in Ref.~\cite{Bulsara} where the AESR was first observed.
Namely, we consider a chain of overdamped, coupled, nonlinear 
(double--well) oscillators

\begin{eqnarray}\label{aesr:coupled}
\dot x_n &=& k x_n - k^\prime x_n^3
+\varepsilon(x_{n-1} - 2x_n + x_{n+1}) 
\nonumber\\
&& + A(\sin\omega t +\phi) + \eta_n(t)\,,
\end{eqnarray}

\noindent where $k=2.1078$, $k^\prime=1.4706$, $A=1.3039$ (these
are the values used in Ref.~\cite{Bulsara}), $\omega=2\pi$, 
$\phi$ is the initial phase of the signal, and
$\eta_n(t)$ are GWNs, possibly spatially correlated according to
\eqref{aesr:etacorrelated}.

If the system responds to the external
periodic stimulation, the central oscillator switches between the two wells
in synchrony with the stimulation. We show results for the ``extreme'' cases
of a~global noise, a local (uncorrelated) noise, and
noises maximally correlated and anticorrelated between the nearest
neighbors. The noises are generated by the algorithm presented in the
preceding Section.
We choose a chain of a modest length of $N=33$, integrate the equations
\eqref{aesr:coupled} numerically using the Heun scheme \cite{Manella}
with a time step $h=1/64$ and analyze the behavior of the central oscillator.
We filter the analog time series 
to generate the time series of $\pm1$, reflecting
which well the oscillator is in. From the power spectrum of
the binary time series we calculate the SNR and average over 64 realizations
of the noises and the initial phases of the signal. Results are presented
in Fig.~\ref{aesr:fig-coupled}.

As we can see, for low noises the SNR curves display rather wild
oscillations, but for larger noises the effect is much the same as for
the nondynamical system discussed above: As the correlations decrease,
the SNR maximum shifts towards higher noise levels. The maximal
value, as well as values for large noise levels, are the largest for 
the noise maximally anticorrelated between the
nearest oscillators and the smallest for the global noise. This effect
grows as the coupling strength increases. 
A more thorough analysis shows that for large noise intensities,
and increase in the SNR is again achieved mainly by lowering the noise
background. Note that in case of the global
noise, the response practically does not change with the coupling strength. 
In this case, all the oscillators receive identical inputs
and there is no AESR, but only the ordinary SR, exactly as in the 
nondynamical system.

To understand the mechanism that is responsible for the AESR in this case,
observe that the oscillators exchange energy via the elastic 
coupling. We calculate the change in the elastic energy between two
neighboring oscillators that occurs during a time interval $h\gtrsim0$,
$\Delta E=E_n(t+h)-E_n(t)$, where 
$E_n(t)=\frac{1}{2}\varepsilon\left(x_{n-1}(t)-x_n(t)\right)^2$.
To the lowest order in $h$, $x_n(t+h)\simeq x_n(t) + h\varphi_n(t)$,
where $\varphi$ has contributions from the elastic interactions, the nonlinear
part of the potential, the external signal, and the noise. If either
the coupling constant $\varepsilon$ is large, or the oscillator happens to
be in the vicinity of the barrier between the wells, the nonlinear part may 
be neglected. Straightforward calculations show that the noises contribute
to the expectation value $\left\langle\Delta E\right\rangle$ a term equal 
to $\varepsilon\sigma^2(1-c_1)$. We can see that noises with negative
correlations maximize this contribution: Noises with negative correlations
tend to pull the neighboring oscillators in the opposite directions, thus
maximizing the energy transfer between the oscillators and providing one
of the oscillators with the extra energy needed to cross the barrier. 
This simple argument
explains why the AESR grows when the coupling strength increases, why 
the AESR is larger for anticorrelated noises, and why the system with 
a~global noise, corresponding to 
$\left\langle\eta_n\eta_{n\pm1}\right\rangle=
\left\langle\eta_n^2\right\rangle$, does not display the AESR. Note that 
if the coupling were repulsive, the situation would be the opposite. 

The effects that the correlations have on the spatiotemporal synchronization
of the system \eqref{aesr:coupled} will be discussed separately.

\section{Discussion}

We have discussed the AESR in arrays of nondynamical threshold
elements. We have not observed any saturation of the SNR curves that was 
reported previously in Ref.~\cite{Collins}. However, the input signal
in that reference was aperiodic and the output SNR is not a natural measure
for such signals \cite{ASR}. This problem has been already 
discussed~\cite{Noest}. We have shown in the present paper that for arrays 
of nondynamical elements, noises with
negative spatial correlations lead to an enhancement of the AESR. In case
of positive spatial correlations the AESR is weaker than for the independent
noises, and arrays with a global noise do not display the AESR, but only
the ordinary SR. This happens because detectors with negative correlations
tend to correct each other's mistakes, while positively correlated
detectors tend to repeat the mistakes. As a result, negatively correlated
detectors better resolve the shape of the incoming signal. The mechanism
of enhancing the SNR relies on lowering the noise background, not on
elevating the signal peak. Note that we
have analyzed these facts for spatial correlations between the nearest neighbors 
only. We expect that noises with long-ranging negative correlations 
would resolve the shape of the incoming signal even better; this point will
be discussed elsewhere. It should be noted, though, that long-ranging 
correlations are more costly to generate.

These results are, superficially, in disagreement with those of Ref.~\cite{Bambi},
where it has been claimed that spatial correlations between the noises 
diminished the positive effect of passing the signal through an array
of model neurons. This discrepancy is easily solved: In Ref.~\cite{Bambi},
each neuron was subjected to a superposition of a subthreshold periodic signal,
a local GWN noise, and a global Orstein-Uhlenbeck noise. The local noises were
mutually uncorrelated and the spatial correlations resulted solely from the 
presence of the global noise. It was the strong global noise that was 
responsible  for the deterioration of the output signal. We have observed 
a~similar effect  --- see Fig.~\ref{aesr:fig-addnoise} above and the subsequent 
discussion. The beneficial effects of negative spatial correlations reported 
here result from the correlations between the local noises.

Our results suggest that from a technological point of view, not only
the additive noises should be incorporated into the design of multi-component
signal-detection systems, as was already suggested in Ref.~\cite{Collins},
but also that these noises should have, if possible, negative spatial
correlations to further improve the system's ability to detect weak signals,
even with a weak global noise present.

We have also shown that spatial correlations of the noise act similarly
in the AESR in a coupled system.
While we have discussed this for the specific system \eqref{aesr:coupled}
only, we have shown that the AESR is enhanced by noises with negative spatial
correlations due to the nature of the attracting harmonic interactions 
between the individual oscillators, regardless of the properties of
the nonlinear part, provided the nonlinear part admits the conventional SR.
For such interactions, positive spatial correlations of the noise reduce
the AESR and a global noise eliminates it altogether, leaving only the
ordinary SR, just like in the case of the AESR in arrays of nondynamical
elements. As harmonic interactions between different particles are
ubiquitous in many physical models, we expect similar phenomena to happen
in a variety of situations. This analysis has also some interesting 
consequences for the interpretation of experimental results with interacting 
agents (particles, oscillators, detectors etc.)
and a global noise, like those reported in Ref.~\cite{brain}: Even though
one cannot examine a single agent and has to excite a group of interacting
ones, with only a global noise added to the signal, the SNR response of the 
whole array is the same as that of a single agent. 

Previous research on systems that display the AESR and our present results
let us conclude that the following features appear not to depend on the
details of the dynamics: (i) For periodic subthreshold inputs, the SNR
is systematically enhanced as the size of the array grows; (ii) Negative
spatial correlations between the local noises provide further enhancement 
of the SNR; (iii) The SNR enhancement is mainly due to lowering the noise
background, not due to increasing the signal peaks; (iv) Positive spatial
correlations reduce the enhancement; in particular, a purely global noise
eliminates the array enhancement altogether. The detailed shape of the
SNR curves, their slope, locations of the maxima, depend on particulars of
the system studied and on properties of the input signals.

\end{document}